\documentstyle[prd,aps,epsf]{revtex}
\tighten
\begin{document}
\draft

\preprint{IC preprint number}

\twocolumn[\hsize\textwidth\columnwidth\hsize\csname @twocolumnfalse\endcsname

\title{A Re-examination of Quenches in $^{4}$He}
\author{G.\ Karra and R.\ J.\ Rivers}
\address{Blackett Laboratory, Imperial College, London SW7 2BZ}
\date{\today}
\maketitle

\begin{abstract}
In the light of recent difficulties in observing vortices in
quenches  of liquid $^{4}$He to its superfluid state we re-examine
the Zurek scenario for their production.  We find that the standard
prediction is suspect because of a confusion over correlation lengths.
\end{abstract}

\pacs{PACS Numbers : 11.27.+d, 05.70.Fh, 11.10.Wx, 67.40.Vs}

\vskip2pc]
%\section*{}

In thermal equilibrium the behaviour of simple systems experiencing
a continuous phase transition is generic, as manifest in the utility
of Landau-Ginzburg theory.  Is this also true dynamically? The
relevance of this question is that the early universe proceeded
through a sequence of phase transitions whose consequences are
directly observable, but whose detailed dynamics is unknown. 
Although it is difficult to measure an order parameter as it
changes, many transitions generate topological charge or topological
defects which can be detected.  Motivated in part by Kibble's
mechanism for the formation of cosmic strings in the early universe\cite{kibble1,kibble2},
Zurek suggested\cite{zurek1} that we measure the density of vortices produced
during a pressure quench of liquid $^{4}He$ into its superfluid
state, as well as the variance of superflow velocity.

The scenario, as proposed by Zurek, is very simple. 
It is exemplified by assuming that the dynamics
of the transition can be derived from an explicitly time-dependent
Landau-Ginzberg free energy of the form
\begin{equation}
F(t) = \int d^{3}x\,\,\bigg(\frac{-\hbar^{2}}{2m}|\nabla\phi |^{2}
+\alpha (t)|\phi |^{2} + \frac{1}{4}\beta |\phi |^{4}\bigg).
\label{F}
\end{equation}
In (\ref{F}) $\phi = (\phi_{1} + i\phi_{2})/\sqrt{2}$ is the
complex order-parameter field, whose magnitude determines the
superfluid density. We identify $\alpha (t)$ as an externally driven
time-dependent chemical potential.  In equilibrium at temperature
$T$, in a mean field approximation $\alpha (T)$ takes the form
$\alpha (T) = \alpha_{0}\epsilon (T_{c})$, where $\epsilon = (T/T_{c}
-1)$ measures the critical temperature
$T_{c}$ relative to $T$. 
In a pressure quench at approximately constant $T$, $T_{c}$ will vary with time $t$, and we
assume that $\epsilon$ can be written as
\begin{equation}
\epsilon (t) = \epsilon_{0} - \frac{t}{\tau_{Q}}\theta (t)
\end{equation}
for $-\infty < t < \tau_{Q}(1 + \epsilon_{0})$, after which
$\epsilon (t) = -1$.  $\epsilon_{0} = (T/T^{in}_{c}
-1)$  measures the original critical temperature $T^{in}_{c}$ against the
temperature $T$ at which the quench takes place, and $\tau_{Q}$
defines the quench rate.  The quench begins at time $t = 0$ and the
transition from the normal to the superfluid phase begins at time $t
= \epsilon_{0}\tau_{Q}$.

With $\xi_{0}^{2} = \hbar^{2}/2m\alpha_{0}$ and $\tau_{0} = \hbar
/\alpha_{0}$ setting the fundamental distance and time scales, the
equilibrium correlation length $\xi (\Delta t)$ and the relaxation
time $\tau (\Delta t)$ diverge at the relative time $\Delta t = t
-\epsilon_{0}\tau_{Q} = 0$ as 
\begin{eqnarray}
\xi (\Delta t) &=& \xi_{0}\bigg(\frac{\Delta t}{\tau_{Q}}\bigg)^{-1/2},
\nonumber
\\
\tau(\Delta t) &=& \tau_{0}\bigg(\frac{\Delta t}{\tau_{Q}}\bigg)^{-1}.
\end{eqnarray}
As we approach the transition, eventually the relaxation time will
be so long that the system will not be able to keep up with the
temperature change.  We estimate the time $t_{Z}$ (and the relative
time $\Delta t_{Z} = t_{Z} - \epsilon_{0}\tau_{Q}$) at which the change from
equilibrium to non-equilibrium behaviour occurs by identifying $\tau(\Delta
t_{Z})$ with $-\Delta t_{Z}$ {\it i.e.}
$-\Delta t_{Z} = \sqrt{t_{0}\tau_{Q}}$.  After this time it is
assumed that the relaxation time is so long  that the field correlation
length $\xi_{Z} = \xi (\Delta t_{Z}) =
\xi_{0}(\tau_{Q}/\tau_{0})^{1/4}$ is more or less frozen in until
the system is again changing slowly, at time $\Delta t \approx +\Delta
t_{Z}$.

The correlation length of the field can only be measured indirectly.
One of Zurek's proposals, as yet unfulfilled, is to measure the
variance in the superflow in an annulus after a quench.  Since
superflow velocity is proportional to the gradient of the
field phase $\theta$, a random walk in phase would suggest that the
measurable $(\Delta\theta )^{2}$ along a perimeter of length $L$ has the form
\begin{equation}
(\Delta\theta )^{2} = O\bigg(\frac{L}{\xi_{var}}\bigg),
\label{var}
\end{equation}
where $\xi_{var}$ measures the effective phase-winding length.
If, as Zurek does, we assume that $\xi_{var}\approx\xi_{Z}$, then $(\Delta\theta
)^{2}$ is large enough to be observed.

A more accessible experiment is to measure the density of vortices at
their formation.   If the {\it initial}
density of vortices, the defects of $^{4}He$, is $n_{def}$, with separation $\xi_{def}$, then
\begin{equation}
n_{def} = O\bigg(\frac{1}{\xi_{def}^{2}}\bigg). 
\end{equation}
Zurek makes the assumption that 
$\xi_{def}\approx\xi_{Z}$ whereby
\begin{equation}
n_{def} = O\bigg(\frac{1}{\xi_{Z}^{2}}\bigg) = 
O\bigg(\frac{1}{\xi_{0}^{2}}\sqrt{\frac{\tau_{Q}}{\tau_{0}}}\bigg).
\label{nZ}
\end{equation}
Since $\xi_{0}$ also measures cold vortex thickness, $\tau_{Q}\gg
\tau_{0}$ corresponds to a measurably large number of widely
separated vortices.  The details do not concern us here, but one
experiment\cite{lancaster} on $^{4}He$, for which $\tau_{Q}/\tau_{0} = O(10^{10})$, 
shows fair agreement with (\ref{nZ}), while
a second experiment (as yet unpublished) with comparable $\tau_{Q}$ finds no vortices, within
its accuracy.  Because of this, it is timely to reexamine this
picture, with its implicit assumption of a domain structure
characterised by a single length.

In this brief note we shall argue that the predictions (\ref{nZ})
and  (\ref{var}) are unreliable because we cannot identify
$\xi_{var}$ and $\xi_{def}$ with $\xi_{Z}$, even though we may
seem to have a single-scale theory. To see this we need an explicit
model, although our conclusions are more general.  Firstly, we
accept that correlation lengths, by whatever definition, are frozen
in during the time interval $-\Delta t_{Z}\leq\Delta t\leq 0$, and
for some time thereafter.  In the
first instance it is then sufficient to perform all our calculations
at  $ \Delta t = 0$.  

At this time
the {\it equal-time} correlation function is diagonal,
\begin{equation}
\langle \phi_{a}({\bf r})\phi_{b}({\bf 0})\rangle = \delta_{ab}G(r).
\end{equation}
Suppose that, as Zurek suggested,
\begin{equation}
G(r)\sim e^{-r/\xi_{Z}}
\label{GZ}
\end{equation}
for $r$ large compared to $\xi_{Z}$ (but not necessarily
asymptotically large).  
To go further we assume that, at this time, field fluctuations are
approximately Gaussian.  An immediate consequence is that the  field
phases are correlated as
\begin{equation}
\langle e^{i\theta({\bf r})}\,e^{-i\theta({\bf 0})}\rangle \approx
\frac{\pi}{4}f(r) \sim e^{-r/\xi_{Z}}
\end{equation}
for large enough $r$ that $f(r)$ is small, where $f(r) = G(r)/G(0)$ is the
normalised correlation function, and $\phi_{a}({\bf r}) = |\phi_{a}({\bf
r})|\,e^{i\theta ({\bf r})}$. Although this seems to support a
simple domain picture, in which field phases are correlated on the
same scale as fields,
the relationship between the unobservable $\xi_{Z}$ and the observable $\xi_{def}$
and $\xi_{var}$ is complicated.  

To begin first with the defect
density, the core of every vortex is a line zero of the complex
field $\phi$.  The converse is not true since zeroes occur on
all scales.  However, it may be valid\cite{popov} to count vortices by counting
line zeroes of an appropriately coarse-grained field, in which
structure on a scale smaller than $\xi_{0}$, the classical vortex size, is ignored.
That is, we ignore vortices within vortices.  This is, indeed, the
basis of the numerous numerical simulations\cite{tanmay,vilenkin} of vortex
networks built from Gaussian fluctuations. 
However, for it to be valid, the result must be insensitive to the
details of the cutoff.  Only then is the estimate reliable.
For the moment, we put in a cutoff $l = O(\xi_{0})$ by hand, as
\begin{equation}
G(r) = \int d \! \! \! / ^3 k\, e^{i{\bf k}.{\bf x}}G(k)\,e^{-k^{2}l^{2}},
\end{equation}
enabling $f(r)$ to be defined.
We stress that if the {\it long-distance} correlation length
$\xi_{Z}$ describes the asymptotic behaviour then it depends only on the position of the
nearest singularity of $G(k)$ in the complex k-plane, {\it
independent} of $l$.  Even if it only characterises intermediate
lengths the dependence on $l$ will be weak.

This is not the case for the line-zero density $n_{zero}$, depending on
the {\it short-distance} behaviour of $G(r)$,
\begin{equation}
n_{zero} = \frac{1}{2\pi\xi_{zero}^{2}} = \frac{-1}{2\pi}f''(0),
\label{ndef}
\end{equation}
the ratio of fourth to second moments of $G(k)\,e^{-k^{2}l^{2}}$. 
$\xi_{zero}$ measures the separation of line zeroes. Only if 
$\partial n_{zero}/\partial l$ is small at $l = \xi_{0}$ can we identify $n_{zero}$
with a meaningful vortex density, and $\xi_{zero}$ with $\xi_{def}$. 

Finally, using the tools of Halperin\cite{halperin} and Mazenko {\it et
al.}\cite{maz}
the final correlation length
$\xi_{var}$ can be shown to be given by
\begin{equation}
\frac{1}{\xi_{var}} = c\int_{0}^{\infty}dr\,\frac{f'(r)^{2}}{1-f(r)^{2}}.
\label{step}
\end{equation}
where $c= O(1)$.
>From (\ref{ndef}) we see that the integrand of (\ref{step}) is
$O(\xi_{zero}^{-2})$ for small $r$, and falls off like
$exp(-2r/\xi_{Z})$ for large enough $r$.  If $\xi_{zero} = O(\xi_{Z})$ then
so is $\xi_{var}$, but otherwise it lies between the two.

Now that we have established the principle that the observable
correlation lengths use different attributes of the power spectrum
$G(k)\,e^{-k^{2}l^{2}}$, we demonstrate it with a concrete example,
motivated by Zurek's later numerical\cite{zurek2} simulations with the time-dependent
Landau-Ginzburg equation.   

That is, for $\Delta t\approx 0$ we assume a linear response
\begin{equation}
\frac{1}{\Gamma}\frac{\partial\phi_{a}}{\partial t} = -\frac{\delta
F}{\delta\phi_{a}} + \eta_{a},
\label{tdlg}
\end{equation}
where $\eta_{a}$ is Gaussian noise. We go further and assume that, for early times, prior
to the symmetry breaking at least,
the self-interaction term can be neglected ($\beta =0$).  This both
preserves Gaussian field fluctuations and leads to $\xi_{Z}$ arising
in a natural way, as we shall see. [A similar approach to
relativistic quantum
field theory\cite{ray} permits a comparison with Kibble's
predictions]. All
correlation lengths depend only on the renormalised $f(r)$, and the
strength of the noise is immaterial. Writing the resultant equation
in time and space units $\tau_{0}$ and $\xi_{0}$ as
\begin{equation}
{\dot \phi}_{a}({\bf k},t) = -\bigg[k^{2} + \bigg({\epsilon_{0} -
\frac{t}{\tau_{Q}}\theta (t)\bigg)\bigg]\phi_{a}(\bf k},t)
+\tau_{0}\eta_{a}({\bf k},t) 
\end{equation}
gives the solution, at $\Delta t =0$, $\phi_{a}({\bf k})=$
\begin{equation}
 =\tau_{0}\int_{-\infty}^{\epsilon_{0}\tau_{Q}}dt
exp\bigg[-\int_{t}^{\epsilon_{0}\tau_{Q}}dt'\bigg[k^{2} + \bigg(\epsilon_{0} -
\frac{t'}{\tau_{Q}}\theta (t)\bigg)\bigg]\eta_{a}({\bf k}).
\end{equation}
The resulting un-normalised correlation function has power
\begin{equation}
G(k) = \int_{-\infty}^{\epsilon_{0}\tau_{Q}}dt
\exp\bigg[-2\int_{t}^{\epsilon_{0}\tau_{Q}}dt'\bigg[k^{2} + \bigg(\epsilon_{0} -
\frac{t'}{\tau_{Q}}\theta (t')\bigg)\bigg]
\label{Gk}
\end{equation}
For a typical quench in $^{4}He$, $\epsilon_{0}\sim 10^{-2} -
10^{-3}$ is very small, but $\tau_{Q}\sim 10^{10}$ is so large that
$\epsilon_{0}\tau_{Q},\,\,\epsilon_{0}^{2}\tau_{Q}\gg 1$. 
$G(k)$ can then be approximated as
\begin{equation}
G(k)=e^{\tau_{Q}k^{4}}\int_{\tau_{Q}k^{2}}^{\tau_{Q}
(\epsilon_{0} + k^{2})}dt\,e^{-t^{2}/\tau_{Q}}.
%\propto e^{\tau_{Q}k^{4}}\,\mbox{erfc}(\sqrt{\tau_{Q}}k^{2})
\label{Gtdlg}
\end{equation}
independent of $\epsilon_{0}$.

On Fourier transforming $G(k)$ of (\ref{Gtdlg}) we find
\begin{equation}
G(r)\propto \int_{0}^{\infty}ds\,s\,e^{-4s^{4}}\,e^{-sr}\,\cos (sr)
\end{equation}  
This does nor give an {\it asymptotic} fall-off of
the form (\ref{GZ}). In fact, for large $r$, $G(r)\propto
\exp(-O((r/\xi_{Z})^{4/3}))$.  
Nonetheless, numerically, it is remarkably well
represented by (\ref{GZ}), with coefficient unity in the exponent,
for $r$ being a few multiples of
$\xi_{Z}$, for reasons that are not clear to us.  In that sense
Zurek's prediction for a correlation length of the form 
$\xi_{Z} = \xi_{0}(\tau_{Q}/\tau_{0})^{1/4}$ is robust.
 
However, it is equally easy to determine the density of line zeroes
at this time. In evaluating
\begin{equation}
\xi_{zero}^{2} = \int_{0}^{\infty}dk\,k^{2}\,e^{-k^{2}l^{2}}G(k)\bigg/
\int_{0}^{\infty}dk\,k^{4}\,e^{-k^{2}l^{2}}G(k)
\label{xi0}
\end{equation}  
we substitute for $G(k)$ from (\ref{Gk}) and perform the $k$
integration first.  For small $\epsilon_{0}$, very large $\tau_{Q}$,
and $l = O(1)$, in dimensionless units the dominant contribution is
from $t\approx\epsilon_{0}\tau_{Q}$. On neglecting terms relatively 
$O(e^{-\epsilon_{0}^{2}\tau_{Q}})$ we find
\begin{equation}
\xi_{zero}^{2}\propto
\int_{0}^{\infty}dt\frac{e^{-t^{2}/\tau_{Q}}}{[t +l^{2}/2]^{3/2}}\bigg/
\int_{0}^{\infty}dt\frac{e^{-t^{2}/\tau_{Q}}}{[t
+l^{2}/2]^{5/2}}\approx O(l^{2}),
\label{xidef0}
\end{equation} 
independent of $\epsilon_{0}$.
The details are immaterial.  Firstly, since $\xi_{zero}\ll\xi_{Z}$ the
frozen correlation length of the field does not set the scale at
which line zeroes appear.  If
$\tau_{0} = 10^{10}$ we have $O(10^{5})$ line zeroes per correlation 
area.
Equally importantly, we have a situation in which the
density of line zeroes depends entirely on the scale at which we look.  If we
look at half the scale ({it i.e.} half a vortex thickness) we see twice as many.  
It is incorrect to
identify the length scale $\xi_{zero}$ with
$\xi_{def}$ since such line zeroes cannot be
understood as vortices.  

This is not surprising.  Although the field correlation length
$\xi_{Z}$ may
have frozen in, the symmetry breaking has not been effected. 
The field magnitude has yet to grow to its equilibrium value
\begin{equation}
\langle |\phi |^{2}\rangle = \beta/\alpha_{0} .
\label{eq}
\end{equation} 
To see how this happens, 
we continue to use (\ref{tdlg}) for times $t >
\epsilon_{0}\tau_{Q}$.  Then,
as the unfreezing occurs, long wavelength modes with $k^{2} < t/\tau_Q -
\epsilon_{0}$ grow exponentially.  Provided
$\epsilon_{0}^{2}\tau_{Q}\gg 1$ they soon begin to dominate the
correlation functions.  Let 
\begin{equation}
G_{n}(\Delta t ) = \int_{0}^{\infty}dk\,k^{n}\,G(k,\Delta t)
\end{equation}
be the moments of
$G(k,\Delta t)$,
now of the form $G(k,\Delta t)=$ 
\begin{equation}
\int_{-\infty}^{\epsilon_{0}\tau_{Q} +\Delta t}dt'
\exp\bigg[-2\int_{t'}^{\epsilon_{0}\tau_{Q} +\Delta t }dt''\bigg[k^{2} + \bigg(\epsilon_{0} -
\frac{t''}{\tau_{Q}}\theta (t'')\bigg)\bigg].
\label{Gkt}
\end{equation}
We find that
\begin{equation}
G_{n}(\Delta t = p\Delta t_{Z})\approx\frac{I_{n}}{2^{n + 1/2}}\,e^{p^{2} }
\int_{0}^{\infty}dt'\,\frac{e^{-(t'-p\sqrt{\tau_{Q}})^{2}/\tau_{Q}}}{[t'
+l^{2}/2]^{n +1/2}},
\label{Gt}
\end{equation}
where we measure the time $\Delta t$ in units of $\Delta t_{Z} = \sqrt{\tau_{Q}}$
from $t = \epsilon_{0}\tau_{Q}$ and $I_{n} = \int_{0}dk k^{2n}\, e^{-k^{2}}$.
If the linear equation (\ref{Gt}) were valid for large $p$ ({\it
i.e.} large $\Delta t$) then the integral is dominated by the
saddle-point at $t'=p\sqrt{\tau_{Q}}$, to give a separation of line
zeroes $\xi_{zero}(\Delta t )$ of the form
\begin{equation}
\xi_{zero}^{2}(\Delta t = p\Delta t_{Z}) = \frac{G_{1}(p\Delta t_{Z})}{G_{2}(p\Delta t_{Z})}
\approx\frac{4p}{3}\xi_{Z}^{2},
\label{falseZ}
\end{equation}
{\it independent} of the cutoff $l$.
That is, because of the transfer of
power to long wavelengths,  line zeroes become widely separated, and
$\xi_{zero}$ does begin to measure vortices, and can be identified
with $\xi_{def}$. 
Thus, if the order parameter is large enough that it takes a long
time (in units of $\Delta t_{Z}$) for the field to populate its
ground states at (\ref{eq}) we would recover the Zurek result
(\ref{nZ}) as an order of magnitude result from (\ref{falseZ}), 
but for entirely different reasons. 

Whether we have time enough depends on the self-coupling $\beta$,
which determines when the linear approximation fails.
For smaller times the integrand gets an increasingly large
contribution from the ultraviolet cutoff dependent lower endpoint, and we recover
(\ref{xidef0}).
The multiple $p_{n}$ of $\Delta t$, at which the exponential modes begin
to dominate in the moments of $G_{n}$, can be determined 
by comparing the relative strengths of the contributions from the
saddlepoint and the endpoint in (\ref{Gt}). For the former to
dominate the latter requires 
\begin{equation}
\frac{e^{p_{n}^{2}}}{p_{n}^{n+1/2}}> 
\frac{(n-1/2)}{\sqrt{\pi}}\bigg(\frac{2\sqrt{\tau_{Q}}}{l^{2}}\bigg)^{n-1/2}.
\end{equation}
We see that it takes longer for the long wavelength modes to
dominate the derivative $G_{2}$ than the order parameter $G_{1}$.
For $\tau_{Q} = O(10^{10})$ and $l\approx 1$, $p_{1}$, satisfying
$e^{p_{1}^{2}}/p_{1}^{3/2}> \tau_{Q}^{1/4}/\sqrt{2\pi}$,
must be
somewhat larger than $2$.  Similarly $p_{2}$, satisfying
$e^{p_{2}^{2}}/p_{2}^{5/2}> 3\sqrt{2}\tau_{Q}^{3/4}/\sqrt{\pi}$
is larger still, somewhat greater than 4.  We stress that these are
lower bounds.

On the other hand, at an absolute maximum, the correlation function
must stop its exponential growth at $\Delta t = \Delta t_{sp} = q\Delta
t_{Z}$, for some multiple $q$, when $\langle |\phi |^{2}\rangle$,
proportional to $G_{1}$, satisfies (\ref{eq}).  Let us suppose that
the effect of the backreaction that stops the growth initially
freezes in any defects.  This then is our prospective starting point for identifying
and counting vortices.  
To determine $q$, or $\Delta t_{sp}$, we need to know the strength of the noise.  In
our dimensionless units we have
\begin{equation}
\langle\eta_{a}({\bf k}, t)\eta_{b}(-{\bf k}', t')\rangle
= \delta_{ab}(2k_{B}T)(\xi_{0}^{3}/\alpha_{0})\delta (t-t')
\delta \! \! \! / ^3 ({\bf k} - {\bf k}').
\end{equation}
As a result, (\ref{eq}) is satisfied when
\begin{equation}
G_{1} (q\Delta t_{Z})= \pi^{2}\alpha_{0}^{2}\xi_{0}^{3}/\beta k_{B}T \approx 10^{2},
\label{Gmax}
\end{equation}
for superfluid $^{4}He$ at $2 K$.
In order that $G_{1}$ is dominated by the exponentially growing
modes, so as to recover Zurek's result in the form (\ref{falseZ}), we must have $q>p_{1},
p_{2}$ where, from (\ref{Gmax})
\begin{equation}
\frac{e^{q^{2}}}{q^{3/2}}< 400\tau_{Q}^{1/4},
\label{tsp}
\end{equation}
say.  In fact, although we need definite figures to make comparisons,
a coefficient like that in (\ref{tsp}) is not entirely believable,
assuming as it does that the backreaction is effectively
instantaneous. However, it is probably good enough to argue that, 
for $\tau_{Q} = O(10^{10})$, $q$ lies {\it between}
$p_{1}$ and $p_{2}$.  That is, field growth must stop before $G_{2}$
is either well defined, or able to give the result (\ref{nZ}).  We
need an impossibly fast quench, with $\tau_{Q} = O(10^{4} - 10^{5})$, for $q$
to be large enough.

What this means for the current $^{4}He$ experiments is unclear. 
Although line zeroes will be separated by more than a classical
vortex thickness they will still show short-range structure and are
only precursors of vortices.  In fact, with $G_{1}$ insensitive to
cutoff $l$, but $G_{2}$ sensitive, $\xi_{zero}$ is arguably {\it more} sensitive to scale than in
(\ref{xi0}), where there was partial cancellation of $l$ dependence
between numerator and denominator. As for $\xi_{var}$ the
calculations are much murkier and we have not attempted
to evaluate it.
Much more detailed numerical modelling in {\it three} dimensions is needed before we can draw
any quantitative conclusions. In less than three dimensions
(particularly in one dimension\cite{dziarmaga}) the
saddle point dominates more easily over the endpoint, which measures the
strength of the ultraviolet singularities.  

Even when our
approximations are correct, it is not clear that the density of vortices
at their first production is the relevant extrapolation for
experiment. The interaction between field modes will tend to
redistribute the power back to shorter wavelengths in the
short-term, and the retarded nature of $G(r)$ after $\Delta t_{sp}$
will impose oscillations that may have consequences, as well as the
possible incorporation of phonon modes.

However, we believe that we have presented a convincing case that,
because line-zero density (and phase variance) are based on
different attributes of the power $G(k)$ from the effective
long-range correlation length $\xi_{Z}$, we should not make the
inference (\ref{nZ}) directly. Only if there is time enough before
the field populates the ground states do we recover the Zurek
result, although for different reasons. More details will be given
elsewhere\cite{algray}.

We thank Alasdair Gill for fruitful discussions.
G.K. would like to thank the Greek State Scholarship Foundation (I.K.Y.) for
financial support. This work is the result of a network supported by the European
Science Foundation, an association of 62 major national funding agencies.

\end{document}